\documentclass[aps,twocolumn,prl,nobibnotes,nofootinbib,showpacs]{revtex4}%
\usepackage{graphicx}
\usepackage{amssymb}
\usepackage{color}%
\usepackage{amsmath}%
\usepackage{amsfonts}
\begin{document}
\title{Dipole blockade through Rydberg F\"{o}rster resonance energy transfer}
\author{Thibault Vogt}
\email{thibault.vogt@lac.u-psud.fr}
\author{Matthieu Viteau}
\author{Jianming Zhao,\footnote{Visitor from College of 	Physics and Electronics Engineering, Shanxi University, China. } Amodsen Chotia}
\author{Daniel Comparat}
\author{Pierre Pillet}
\affiliation{Laboratoire Aim\'{e} Cotton,\footnote{Laboratoire Aim\'{e} Cotton is
associated to Universit\'{e} Paris-Sud and belongs to F\'{e}d\'{e}ration de
Recherche Lumi\`{e}re Mati\`{e}re (LUMAT).} CNRS, B\^{a}t. 505, Campus
d'Orsay, 91405 Orsay, France}
\date{15 February 2006}

\begin{abstract}
High resolution laser excitation of $np$ Rydberg states of cesium atoms shows
a dipole blockade at F\"{o}rster resonances corresponding to the resonant
dipole-dipole energy transfer of the $np+np\longrightarrow ns+(n+1)s $
reaction. The dipole-dipole interaction can be tuned on and off by the Stark
effect, and such a process observed for relatively low $n$ ($25-41$) is
promising for quantum gate devices. Both Penning ionization and saturation in
the laser excitation can limit the range of observation of the dipole blockade.

\end{abstract}
\maketitle

Rydberg atoms have long been known to possess huge electric dipole moments
leading to exaggerated collisional properties of room temperature atoms, in
particular, large cross sections and long interaction times \cite{Gallagher}%
.\ These properties have stimulated great interest
in the possibility of controlling the strong long-range interactions between cold
atoms \cite{Mourachko,Anderson,Fioretti}, which could be particularly exciting for
quantum information applications \cite{Jaksch,Lukin}. One interesting process
is the possibility of dipole blockade in the Rydberg excitation of atoms, due
to the dipole-dipole interaction shifting the Rydberg energy from its isolated
atomic value. If the volume of the laser excitation is small enough, no second
atom can be excited after the Rydberg excitation of the first one, producing
an atomic ensemble in a singly excited collective state. The use of the dipole
blockade of the excitation has been proposed as a very efficient realization
of a scalable\ quantum logic gate \cite{Lukin}. If we consider the laser
high-resolution-excitation of a large ensemble of atoms from the ground state
or from a low-excited state towards a Rydberg state, the dipole-dipole
interaction between Rydberg atoms should lead to a limitation of the number of
excited atoms versus the initial density of atoms.\ A partial, or local,
blockade of the excitation is thus expected.\ The first excited Rydberg atoms
shift the resonance for their non-excited neighbors and prevent their
excitation with a narrow-bandwidth laser.

Up to now no evidence of a dipole blockade has been demonstrated.\ In zero
electric field, the atomic Rydberg state do not have permanent dipole moments
and no dipole blockade is expected. Second order dipole-dipole or van der
Waals coupling between Rydberg atoms can occur, and a suppression of the
excitation corresponding to a partial, or local, blockade has been reported in
excitation of high Rydberg states ($n\sim70-80$) using a pulse amplified
single mode laser \cite{Tong}. Cw excitations have also been performed showing
the supression of the excitation \cite{Singer} and studying the sub-Poissonian
atom counting statistics \cite{RaithelPRL05}.\ The authors notice that for
long-duration excitations, appearance of ions can lift the blockade.\ For a
broad-band excitation, the suppression of the excitation is not
expected.\ Rather a band of levels can be excited with a density-dependent
broadening, which has been probed using micro-wave transitions
\cite{Afrousheh}, or laser depumping \cite{MudrichPRL05}. In high electric
field, the Stark effect mixes the Rydberg states, and the Rydberg atoms have a
significant permanent dipole moment. The dipole blockade is expected but the
presence of both dipole-dipole and van der Waals couplings complicate
the evolution of the blockade versus the electric field.\ An alternative is
the use of a configuration of strong dipole-dipole coupling in low electric
field with relatively low Rydberg states, similar to F\"{o}rster resonances
\cite{GallagherPRA82}. The F\"{o}rster (or Fluorescence) Resonance Energy
Transfer (FRET) is commonly used as spectroscopic technique based on
long-range dipole-dipole coupling to estimate small distance and dynamics
between fluorescent molecules \cite{Foerster,Foerster59}.

In this letter, we report the observation of the dipole blockade of the
Rydberg excitation by cw lasers, by taking advantage of a resonant
dipole-dipole coupling of cesium atoms, specifically the process
\begin{equation}
np_{3/2}+np_{3/2}\longrightarrow ns+(n+1)s.\label{eq:restransfer}%
\end{equation}
The dipole-dipole interaction is $W\sim\mu\mu^{\prime}R^{-3}$, where $\mu,$
$\mu^{\prime}\sim n^{2}$ $atomic$ $units$\ are the dipoles corresponding to
the transitions $np_{3/2}\longrightarrow ns$ and $np_{3/2}\longrightarrow
(n+1)s$, respectively, and $R$\ the distance between the two atoms.\ Control
of the dipole-dipole coupling is achieved by using the Stark effect.\ The
resonance is obtained when, by Stark shifting, the level $np_{3/2}$\ is
located midway in the energy diagram between the states, $ns$ and $(n+1)s$
\cite{Mourachko}. At such resonances, also called F\"{o}rster resonances, we
observe a significant decrease of the $p$ excitation, interpreted as a local
dipole blockade.

\begin{figure}
[ptb]
\begin{center}
\includegraphics[
height=1.9435in,
width=3.1648in
]%
{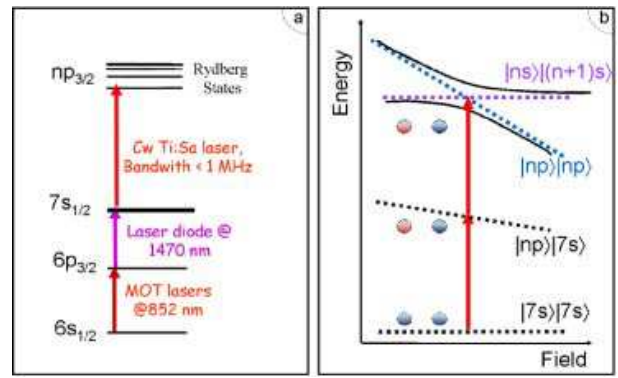}%
\caption{(Color online) Representations (a) of the three-step scheme of
excitation for $np_{3/2}$ Rydberg states, and (b) of the excitation of a pair
of atom at the F\"{o}rster resonance. The dipole-dipole coupling leads to an
avoiding crossing between the energy levels $\left\vert np_{3/2}\right\rangle
\otimes\left\vert np_{3/2}\right\rangle $ and $\left\vert ns\right\rangle
\otimes\left\vert (n+1)s\right\rangle $ of the pair of atoms.}%
\label{fig:levels}%
\end{center}
\end{figure}

The details of the experimental setup have been described in several papers
\cite{Mourachko,Fioretti,MudrichPRL05}.\ But here, three cw
lasers provide a new high resolution multistep scheme of excitation, as depicted in Fig.\ref{fig:levels} (a). The
Rydberg atoms are excited from a cloud of 5$\ 10^{7}$ cesium atoms
(temperature 200 $\mu$K,characteristic radius $\sim$\ 300 $\mu$m, peak and average density 2 10$^{11}$\ and 4 10$^{10}$
cm$^{-3}$, respectively) produced in a standard vapor-loaded magneto-optical
trap (MOT) at residual gas pressure of $10^{-10}$\thinspace mbar. The first
step of the excitation, $6sF=4\rightarrow6p_{3/2}F=5$, is provided by the
trapping lasers of the MOT.\ The density of excited, $6p_{3/2}$, atoms present
in the MOT can be modified by switching off the repumper lasers of the MOT,
with a delay ranging from 0 to 1 ms before applying the other steps of the
excitation.\ We use an infrared diode laser (DL100\ TOPTICA) with a wavelength
of $\lambda_{2}=1.47$ $\mu$m\ in an extended cavity device for the second step
of the excitation, $6p_{3/2}F=5\rightarrow7sF=4$. The bandwidth is of the
order of 100 kHz and the available power is 20 mW.\ The last step of the
excitation, $7sF=4\rightarrow np_{3/2}$ (with $n=25-45$), is provided by a
Titanium:Sapphire (Ti:Sa) laser pumped by an $Ar^{+}$ laser. The wavelength
$\lambda_{3}$ ranges from $770$ to $800$ nm, the bandwidth is $1$ MHz, and the
available power is $400\,$mW. The Ti:Sa laser is switched on during a time,
$\tau=0.3\,\mu$s by means of an acousto-optic modulator. The beams of the
infrared diode laser and of \ the Ti:Sa laser cross with an angle of
67.5\ degrees and are focused into the atomic cloud with waists of 105 and 75
$\mu$m respectively. Their polarizations are both linear
and parallel to the direction of the applied electric field, leading to the
excitation of the magnetic sublevel $\left\vert m\right\vert =1/2$ for
$np_{3/2}$. The spectral resolution, $\Delta\nu$, of the excitation is of the
order of $5-6$ MHz, limited by the lifetime, 56.5 $ns$, of the $7s$\ state and
by the duration and the spectral width of the Ti:Sa laser pulse.\ The magnetic
quadrupole field of the MOT is not switched off during the Rydberg excitation
phase, it contributes for less than 1 MHz to the observed linewidths. The
average intensity ($\sim3$ mW/cm$^{2}$) of the infrared diode laser is chosen
to reach saturation of the transition ($\sim1.4$ mW/cm$^{2}$). At the trap
position, a static electric field and a pulsed high voltage field are applied
by means of a pair of electric field grids spaced by $15.7\,$mm. \ We analyze
selectively the Rydberg states by applying just after the Ti:Sa laser pulse a
field ramp with a rise time of 700 ns, and by detecting the ions with a pair
of microchannel plates.%

\begin{figure}
[ptb]
\begin{center}
\includegraphics[
trim=0.227422in 0.187189in 0.213456in 0.325724in,
height=2.3495in,
width=3.0577in
]%
{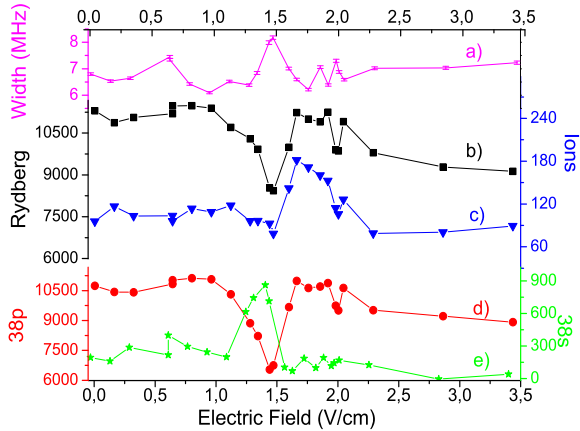}%
\caption{(Color online) Detailed study of the dipole blockade for the
$38p_{3/2}$ state. The Ti:Sa laser is resonant with the transition
$7s\longrightarrow38p_{3/2}$ and its power is kept constant at 8 mW
corresponding to an average intensity of 45\ W/cm$^{2}$.\ Versus the applied
electric field: (a) linewidths, (b) total number of Rydberg atoms, (c) number
of ions, (d) and (e) number of atoms in $38p_{3/2}$ and $38s$, respectively.}%
\label{fig:DipBlockade38p}%
\end{center}
\end{figure}

The experimental procedure consists of Stark spectroscopy of the $np_{3/2}$
states at a repetition rate of 80 Hz for different atomic densities, and for
different Ti:Sa laser intensities.\ During the excitation process a small
number, $\sim100$, of ions less that one percent of the total number of
Rydberg atoms is formed, which we attribute to cold collisions between Rydberg
atoms \cite{LiPRL05}. These ions are simultaneously detected with the Rydberg
atoms. Fig.\ref{fig:DipBlockade38p} shows the analysis of spectra of the
$38p_{3/2}$ state performed at different values of the electric field. At zero
electric field, the number of \ excited Rydberg atoms is about $11\ 000$,
corresponding to an estimated average density of $2 $ $10^{9}$ cm$^{-3}$.\ The
electric field dependence of the width, the total number of atoms excited, and
the number of ions present in the Rydberg sample at the end of the excitation
are shown in Figs.\ref{fig:DipBlockade38p} (a, b, c). First we observe a
minimum in the number of excited atoms at the resonance of the reaction of Eq.
(\ref{eq:restransfer}), at $F=1.46$ V/cm.\ The width of the lines away from
the F\"{o}rster resonance is 6 MHz, the limit of resolution of the multistep
laser excitation. It is increased at the resonance to $8-9$ MHz. At the
resonance, we do not observe any increase in the number of ions. In the high
field wing of the resonance, we can observe a small increase of this number,
up to 200, which corresponds to a Penning ionization process by exciting pairs
of close atoms which are subjected to an attractive force
\cite{LiPRL05,MudrichPRL05}. The number of ions stays small enough to add no
broadening to the line.
Figs.\ref{fig:DipBlockade38p} (d, e) show the selective detection of the
population in the $38p_{3/2}$ state and the population transferred in the
$38s$\ state versus the electric field. The resonance in the $s$\ signal is at
the same position as the minimum of the $p $ excitation, both curves
presenting the same width versus the electric field.%

\begin{figure}
[pb]
\begin{center}
\includegraphics[
trim=0.161823in 0.061527in 0.247469in 0.283970in,
height=1.638in,
width=3.1224in
]%
{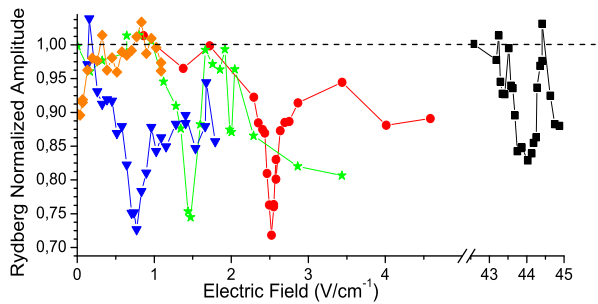}%
\caption{(Color online) Normalized number of excited Rydberg atoms, for the
same experimental conditions as in Fig.\ref{fig:DipBlockade38p}, versus the
applied electric field for different $np_{3/2}$ states: $42p_{3/2}$ (diamonds
symbols), $40p_{3/2}$ (triangles), $38p_{3/2}$ (stars), $36p_{3/2}$ (circles),
and $25p_{3/2}$ (squares). The F\"{o}rster resonances occur at fields of 0.75,
1.46, 2.53, and 44.03 V/cm. The normalisation is performed by considering the
maximum field amplitude, except $n=25$, for which the signal at 43.17 V/cm is
normalized to one.}%
\label{fig:DipBlock42-25p}%
\end{center}
\end{figure}

Fig.\ref{fig:DipBlock42-25p} shows the evolution of the amplitudes of the
$np_{3/2}$ excitation versus the electric field for different principal
quantum numbers $n=42,$ $40,$ $38,$ $36,$ and $25$. Except for $n=42$ , we
observe a minimum corresponding to the resonance of the reaction of Eq.
(\ref{eq:restransfer}). For $n\geq42$, the energy of a
pair of atoms $np_{3/2}+np_{3/2}$ becomes smaller than the one of a pair
$ns+(n+1)s$, and the F\"{o}rster resonance no longer exists. Nevertheless at
$n=42$, the difference between the energy levels of a pair of atoms is only
$-9.5$ MHz, small enough to observe a decrease of the excitation at zero
field.\ The data show a clear dipole blockade of the excitation of up to
$30\%$\ for $n=40,$ $38,$ and $36$. For each $n$, we observe also a small
increase in the width of the spectral line of the $p_{3/2}$ excitation at the
F\"{o}rster resonance. For lower $n$, $\sim25$, a less important dipole
blockade of the excitation is also observed but no real broadening of the
spectral lines.%

\begin{figure}
[ptb]
\begin{center}
\includegraphics[
trim=0.194182in 0.125820in 0.179970in 0.173287in,
height=2.2026in,
width=2.7522in
]%
{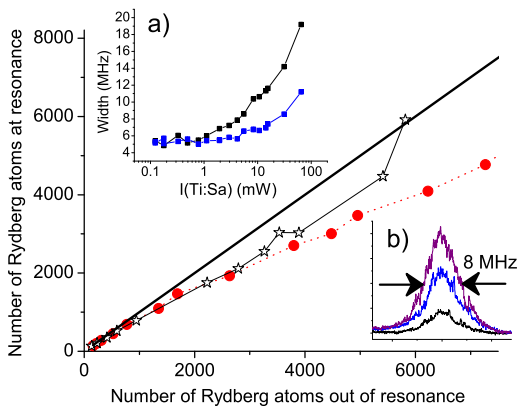}%
\caption{(Color online) Number of $36p_{3/2}$ Rydberg atoms excited at the
F\"{o}rster resonance (2.53\ V/cm) versus the one out of resonance (1.37
V/cm), for two different initial atomic densities of $7s$ state, 8 10$^{9}$
(filled circles) and 2.7\ 10$^{9}$ (empty stars). The efficiency of the
blockade is the gap to the full (with a slope of 1) and broken lines. (a)
Linewidths of laser resonance ($7s\longrightarrow36p_{3/2}$) at two electric
field values: 2.52 (F\"{o}rster resonance: upper curve), 1.37 (lower curve)
V/cm. (b) Resonances for 0.8, 3.0\ and 4.2 mW\ laser power.\ The broadening
varies from 6 to 8 MHz as the laser power.}%
\label{fig:DipBlock36p-IntLaser}%
\end{center}
\end{figure}

One property of the dipole blockade of the excitation is that it should be
dependent on the atomic density and on the laser excitation.
Fig.\ref{fig:DipBlock36p-IntLaser} shows the evolution of the dipole blockade
in the case of the $36p_{3/2}$\ state. We have plotted the number of excited
Rydberg atoms, $36p_{3/2}$, at the F\"{o}rster resonance versus the one out of
resonance obtained for two different average atomic densities corresponding
for the $7s$ state, respectively $\sim8\pm3$ $10^{9}$ (which is used for most
of reported results) and $2.7$ $10^{9}$ cm$^{-3}$. First we see the appearance
of the dipole blockade when the number of Rydberg atoms exceeds 1000. Both
curves present then the same blockade up to a Rydberg excitation of
3500\ atoms.\ Above this point, the evolution of both curves differs.\ When
the number of Rydberg atoms exceeds 5000, the effect of dipole blockade
disappears in the case of an initial $7s$ state density of $2.7$ $10^{9}$
cm$^{-3}$, while its efficiency goes on to increase for a three times larger
density. This result can seem quite surprising and is linked to the saturation
of the excitation. Experimentally, the number of Rydberg atom varies linearly
with the power of the Ti:Sa laser up to 4 mW, corresponding to an intensity of
$22$ W/cm$^{2}$, then the excitation starts to saturate.\ For instance on
Fig.\ref{fig:DipBlock36p-IntLaser}, to obtain 5000 Rydberg atoms, we need five
times more power (6 and 30 mW, respectively) while the atomic density of $7s$
state is only three times smaller.\ We observe a broadening of the resonance
with the laser power (see Fig.\ref{fig:DipBlock36p-IntLaser} (a)) independent
on the atomic density.\ By increasing the laser power from 0.1\ to 10 mW\ we
observe an increase in the efficiency of the dipole blockade.\ If we increase
more the power of the Ti:Sa laser the dipole blockade becomes less efficient.
Indeed, at high laser power the resonance lines are broadened, such that the
laser excitation of pairs of closer atoms become possible, preventing the
dipole blockade.\ The range in density and laser intensity allowing
observation of the dipole blockade is therefore limited, and the dipole
blockade results shown in Figs.\ref{fig:DipBlockade38p} and
\ref{fig:DipBlock42-25p} correspond to the optimum of our experimental
conditions, with a low power of the Ti:Sa laser (linewidth $\sim\Delta\nu$) and a high atomic density. At
this point of this letter, we understand most of the features of the
experiment, the observed dipole blockade seems to indicate that at the
F\"{o}rster resonance no pair of close atoms can be excited at a distance
smaller than $R_{\min}\sim \left(\mu\mu^{\prime}/W \right)^{1/3} \sim \left(  \mu\mu^{\prime}/ \Delta\nu\right)  ^{1/3}$,
which corresponds for $n=38$ to $\sim4$ $\mu$m, in agreement with the measured
Rydberg atom number and average Rydberg density, 8500\ and $1.7$ $10^{9}$
cm$^{-3}$ ($W \sim$ 2.5 MHz by
considering the average density and 5.5 MHz for the peak one).

An unexpected result is that the width of the F\"{o}rster resonance versus the
static electric field (see Fig.1 (b)), which converted to frequency ($50\pm
10$\ MHz in the case of Fig.\ref{fig:DipBlockade38p} (b)) is much larger than
the width of the spectral line (8\ MHz shown in the inset of
Fig.\ref{fig:DipBlock36p-IntLaser} (b)).\ The resolution of this paradox lies
probably in the dynamics of the dipole-dipole coupling during the Rydberg
excitation, where many-body interactions are not negligible when the Rydberg
density increases \cite{Mourachko,Anderson,Akulin99,FrasierPRA99}. The
many-body effects are linked to the migration of the $s$ state ($ns$ or
$(n+1)s$) excitation in the $p$ state environment by excitation exchange due
to dipole-dipole coupling in the frozen Rydberg gas, $i.e.$ by the resonant
processes:

\begin{align}
ns+np_{3/2}  & \longrightarrow np_{3/2}+ns,\nonumber\\
\text{and }(n+1)s+np_{3/2}  & \longrightarrow np_{3/2}%
+(n+1)s.\label{eq:migration}%
\end{align}
The $R^{-3}$ character of the dipole-dipole interaction makes the dipole
blockade mostly result of two-body effects, meaning during the excitation each
atom is essentially sensitive to its closest Rydberg neighbor. Nevertheless,
in the local blockade process a pair of atoms at distance of the order of
$R_{\min}$ can be excited and interact with the energy transfer of Eq.
(\ref{eq:restransfer}) and the many-body effects of Eq. (\ref{eq:migration})
occur. The role of the migration of the $s$ state excitation is to
increase\ the number of $s$ Rydberg atoms in the sample, because a pair of
close Rydberg, $p$, atoms can react several times \cite{Mourachko}.\ Such an
ensemble of $s$ excitation enhances probably the blockade outside F\"{o}rster
resonance, where the $sp\longrightarrow ps$ dipole-dipole interaction is
always resonant, while the reaction of Eq.(\ref{eq:restransfer}) is no longer
resonant and less efficient.\ The broad band of the F\"{o}rster resonance is a
signature for the many-body behavior, while the narrow laser resonances are
ones for two-body interaction.%

To conclude, we have shown clear evidence for a local dipole blockade of the
Rydberg excitation at a F\"{o}rster resonance, controlled via the Stark effect. This result is not only promising for quantum
information but offers a lot of applications. The use
of a second laser scanned around the transition $7s\longrightarrow np_{3/2}$
should permit us to excite pairs of very close atoms with a repulsive or an
attractive mutual force.\ The band of levels due to the Rydberg interactions
can also be probed by microwave excitation \cite{MourachkoPRA,Afrousheh}%
.\ The precise shape of the resonance should be an excellent probe of the
random character of the distribution of the pairs \cite{Akulin06}.\ The Stark
control of the dipole blockade should also allow us to prepare Rydberg atomic
samples with pairs of repulsive or attractive atoms to analyze the formation
of ions and the evolution towards an ultracold plasma \cite{Robinson,LiPRA}.
The observation of the total dipole blockade, meaning the excitation of a
single atom, is still a challenge.\ It implies a volume of the Rydberg sample
with a radius of the order of $R_{\min}$. Here we observe the F\"{o}rster
resonance up to $n=41$, limiting $R_{\min}$ at the value of $4$ $\mu$m. The
choice of another F\"{o}rster configuration with $n$ of the order of 100 or
more increases $R_{\min}$ up to $20$ $\mu$m. A microwave field instead of the
Stark effect could also be used to produce the dipole moments and to control
the dipole blockade \cite{PilletPRA87,LesHouches04}.

This work is in the frame of "Institut francilien de recherche sur les atomes
froids" (IFRAF) and of the European Research and Training Networks COLMOL
(HPRN-CT-2002-00290) and QUACS (HPRN-CT-2002-00309). One of the authors (J.
Z.) is supported by IFRAF. The cw excitation development corresponds to a
preliminary study for the CORYMOL experiment supported by an ANR\ grant
(NT05-2 41884).\ The authors thank D. A. Tate for his helpful participation to preliminary work on this experiment and acknowledge very fruitful discussions with Thomas F.\ Gallagher, Marcel Mudrich, Nassim Zahzam, and Vladimir Akulin.

\bibliographystyle{apsrev}

\begin{thebibliography}{23}
\expandafter\ifx\csname natexlab\endcsname\relax\def\natexlab#1{#1}\fi
\expandafter\ifx\csname bibnamefont\endcsname\relax
  \def\bibnamefont#1{#1}\fi
\expandafter\ifx\csname bibfnamefont\endcsname\relax
  \def\bibfnamefont#1{#1}\fi
\expandafter\ifx\csname citenamefont\endcsname\relax
  \def\citenamefont#1{#1}\fi
\expandafter\ifx\csname url\endcsname\relax
  \def\url#1{\texttt{#1}}\fi
\expandafter\ifx\csname urlprefix\endcsname\relax\def\urlprefix{URL }\fi
\providecommand{\bibinfo}[2]{#2}
\providecommand{\eprint}[2][]{\url{#2}}

\bibitem[{\citenamefont{Gallagher}(1994)}]{Gallagher}
\bibinfo{author}{\bibfnamefont{T.~F.} \bibnamefont{Gallagher}},
  \emph{\bibinfo{title}{Rydberg Atoms}} (\bibinfo{publisher}{Cambridge
  University Press, New York}, \bibinfo{year}{1994}).

\bibitem[{\citenamefont{Mourachko{\textit{ et al.}}}(1998)}]{Mourachko}
\bibinfo{author}{\bibfnamefont{I.}~\bibnamefont{Mourachko{\textit{ et al.}}}},
  \bibinfo{journal}{Phys.\ Rev.\ Lett.} \textbf{\bibinfo{volume}{80}},
  \bibinfo{pages}{253} (\bibinfo{year}{1998}).

\bibitem[{\citenamefont{Anderson{\textit{ et al.}}}(1998)}]{Anderson}
\bibinfo{author}{\bibfnamefont{W.~R.} \bibnamefont{Anderson{\textit{ et
  al.}}}}, \bibinfo{journal}{Phys.\ Rev.\ Lett.} \textbf{\bibinfo{volume}{80}},
  \bibinfo{pages}{249} (\bibinfo{year}{1998}).

\bibitem[{\citenamefont{Fioretti{\textit{ et al.}}}(1999)}]{Fioretti}
\bibinfo{author}{\bibfnamefont{A.}~\bibnamefont{Fioretti{\textit{ et al.}}}},
  \bibinfo{journal}{Phys.\ Rev.\ Lett.} \textbf{\bibinfo{volume}{82}},
  \bibinfo{pages}{1839} (\bibinfo{year}{1999}).

\bibitem[{\citenamefont{Jaksch{\textit{ et al.}}}(2000)}]{Jaksch}
\bibinfo{author}{\bibfnamefont{D.}~\bibnamefont{Jaksch{\textit{ et al.}}}},
  \bibinfo{journal}{Phys.\ Rev.\ Lett.} \textbf{\bibinfo{volume}{85}},
  \bibinfo{pages}{2208} (\bibinfo{year}{2000}).

\bibitem[{\citenamefont{Lukin{\textit{ et al.}}}(2001)}]{Lukin}
\bibinfo{author}{\bibfnamefont{M.~D.} \bibnamefont{Lukin{\textit{ et al.}}}},
  \bibinfo{journal}{Phys.\ Rev.\ Lett.} \textbf{\bibinfo{volume}{87}},
  \bibinfo{pages}{037901} (\bibinfo{year}{2001}).

\bibitem[{\citenamefont{Tong{\textit{ et al.}}}(2004)}]{Tong}
\bibinfo{author}{\bibfnamefont{D.}~\bibnamefont{Tong{\textit{ et al.}}}},
  \bibinfo{journal}{Phys.\ Rev.\ Lett.} \textbf{\bibinfo{volume}{93}},
  \bibinfo{pages}{063001} (\bibinfo{year}{2004}).

\bibitem[{\citenamefont{Singer{\textit{ et al.}}}(2004)}]{Singer}
\bibinfo{author}{\bibfnamefont{K.}~\bibnamefont{Singer{\textit{ et al.}}}},
  \bibinfo{journal}{Phys.\ Rev.\ Lett.} \textbf{\bibinfo{volume}{93}},
  \bibinfo{pages}{163001} (\bibinfo{year}{2004}).

\bibitem[{\citenamefont{Liebish{\textit{ et al.}}}(2005)}]{RaithelPRL05}
\bibinfo{author}{\bibfnamefont{T.} \bibnamefont{CubelLiebisch{\textit{ et al.}}}},
  \bibinfo{journal}{Phys.\ Rev.\ Lett.} \textbf{\bibinfo{volume}{95}},
  \bibinfo{pages}{253002} (\bibinfo{year}{2005}).

\bibitem[{\citenamefont{Afrousheh{\textit{ et al.}}}(2004)}]{Afrousheh}
\bibinfo{author}{\bibfnamefont{K.}~\bibnamefont{Afrousheh{\textit{ et al.}}}},
  \bibinfo{journal}{Phys.\ Rev.\ Lett.} \textbf{\bibinfo{volume}{93}},
  \bibinfo{pages}{233001} (\bibinfo{year}{2004}).

\bibitem[{\citenamefont{Mudrich{\textit{ et al.}}}(2005)}]{MudrichPRL05}
\bibinfo{author}{\bibfnamefont{M.}~\bibnamefont{Mudrich{\textit{ et al.}}}},
  \bibinfo{journal}{Phys.\ Rev.\ Lett.} \textbf{\bibinfo{volume}{95}},
  \bibinfo{pages}{233002} (\bibinfo{year}{2005}).

\bibitem[{\citenamefont{Gallagher{\textit{ et al.}}}(1982)}]{GallagherPRA82}
\bibinfo{author}{\bibfnamefont{T.~F.} \bibnamefont{Gallagher{\textit{ et
  al.}}}}, \bibinfo{journal}{Phys.\ Rev.\ A} \textbf{\bibinfo{volume}{25}},
  \bibinfo{pages}{1905} (\bibinfo{year}{1982}).

\bibitem[{\citenamefont{F\"orster}(1996)}]{Foerster}
\bibinfo{author}{\bibfnamefont{T.}~\bibnamefont{F\"orster}},
  \emph{\bibinfo{title}{\textrm{in} Modern Quantum Chemistry}}
  (\bibinfo{publisher}{Academic Press, New York}, \bibinfo{year}{1996}).

\bibitem[{\citenamefont{F\"orster}(1959)}]{Foerster59}
\bibinfo{author}{\bibfnamefont{T.}~\bibnamefont{F\"orster}},
  \bibinfo{journal}{Discuss.\ Faraday Soc.} \textbf{\bibinfo{volume}{27}},
  \bibinfo{pages}{7} (\bibinfo{year}{1959}).

\bibitem[{\citenamefont{Li et~al.}(2005)\citenamefont{Li, Tanner, and
  Gallagher}}]{LiPRL05}
\bibinfo{author}{\bibfnamefont{W.}~\bibnamefont{Li}},
  \bibinfo{author}{\bibfnamefont{P.~J.} \bibnamefont{Tanner}},
  \bibnamefont{and} \bibinfo{author}{\bibfnamefont{T.~F.}
  \bibnamefont{Gallagher}}, \bibinfo{journal}{Phys.\ Rev.\ Lett.} \textbf{\bibinfo{volume}{94}},
  \bibinfo{pages}{173001} (\bibinfo{year}{2005}).
  
\bibitem[{\citenamefont{Akulin{\textit{ et al.}}}(1999)}]{Akulin99}
\bibinfo{author}{\bibfnamefont{W.~M.} \bibnamefont{Akulin{\textit{ et al.}}}},
  \bibinfo{journal}{Physica D} \textbf{\bibinfo{volume}{131}},
  \bibinfo{pages}{125} (\bibinfo{year}{1999}).

\bibitem[{\citenamefont{Frasier{\textit{ et al.}}}(1999)}]{FrasierPRA99}
\bibinfo{author}{\bibfnamefont{J.~S.} \bibnamefont{Frasier{\textit{ et al.}}}},
  \bibinfo{journal}{Phys.\ Rev.\ A} \textbf{\bibinfo{volume}{59}},
  \bibinfo{pages}{4358} (\bibinfo{year}{1999}).

\bibitem[{\citenamefont{Mourachko{\textit{ et al.}}}(2004)}]{MourachkoPRA}
\bibinfo{author}{\bibfnamefont{I.}~\bibnamefont{Mourachko{\textit{ et al.}}}},
  \bibinfo{journal}{Phys.\ Rev.\ A} \textbf{\bibinfo{volume}{70}},
  \bibinfo{pages}{031401(R)} (\bibinfo{year}{2004}).

\bibitem[{\citenamefont{Akulin}(2006)}]{Akulin06}
\bibinfo{author}{\bibfnamefont{V.~M.} \bibnamefont{Akulin}},
  \emph{\bibinfo{title}{Coherent Dynamics of Complex Quantum Systems}}
  (\bibinfo{publisher}{Springer}, \bibinfo{year}{2006}).

\bibitem[{\citenamefont{Robinson{\textit{ et al.}}}(2000)}]{Robinson}
\bibinfo{author}{\bibfnamefont{M.~P.} \bibnamefont{Robinson{\textit{ et
  al.}}}}, \bibinfo{journal}{Phys.\ Rev.\ Lett.} \textbf{\bibinfo{volume}{85}},
  \bibinfo{pages}{4466} (\bibinfo{year}{2000}).

\bibitem[{\citenamefont{Li{\textit{ et al.}}}(2004)}]{LiPRA}
\bibinfo{author}{\bibfnamefont{W.}~\bibnamefont{Li{\textit{ et al.}}}},
  \bibinfo{journal}{Phys.\ Rev.\ A} \textbf{\bibinfo{volume}{70}},
  \bibinfo{pages}{042713} (\bibinfo{year}{2004}).

\bibitem[{\citenamefont{Pillet{\textit{ et al.}}}(1987)}]{PilletPRA87}
\bibinfo{author}{\bibfnamefont{P.}~\bibnamefont{Pillet{\textit{ et al.}}}},
  \bibinfo{journal}{Phys.\ Rev. A} \textbf{\bibinfo{volume}{36}},
  \bibinfo{pages}{1132} (\bibinfo{year}{1987}).

\bibitem[{\citenamefont{Pillet{\textit{ et al.}}}(2005)}]{LesHouches04}
\bibinfo{author}{\bibfnamefont{P.}~\bibnamefont{Pillet{\textit{ et al.}}}},
  \emph{\bibinfo{title}{in Decoherence, Entanglement and Information Protection
  in Complex Quantum Systems (pages 411-436), edited by V. M. Akulin{\textit{
  et al.}}}}, vol. \bibinfo{volume}{189} (\bibinfo{publisher}{NATO Sciences
  Series II}, \bibinfo{year}{2005}).

\end{thebibliography}

\end{document}